\documentclass[prd,onecolumn,
  tightenlines,nofootinbib,eqsecnum,
]{revtex4}

\usepackage[francais]{babel}

\usepackage{amsmath}
\usepackage{amsfonts}
\usepackage{amssymb}
\usepackage{bm}
\usepackage{hyperref}
\usepackage{mathrsfs}
\usepackage{graphicx}

\usepackage{empheq}

\usepackage{ulem}
\usepackage[usenames]{color}

\definecolor{darkgreen}{rgb}{0,0.5,0}

\hypersetup{
    bookmarks=true,         
    unicode=false,          
    pdftoolbar=true,        
    pdfmenubar=true,        
    pdffitwindow=false,     
    pdfstartview={FitH},    
    pdftitle={My title},    
    pdfauthor={Author},     
    pdfsubject={Subject},   
    pdfcreator={Creator},   
    pdfproducer={Producer}, 
    pdfkeywords={keyword1} {key2} {key3}, 
    pdfnewwindow=true,      
    colorlinks=true,       
    linkcolor=red,          
    citecolor=cyan,        
    filecolor=magenta,      
    urlcolor=darkgreen,           
    linktocpage=true
}

\allowdisplaybreaks

\DeclareSymbolFontAlphabet{\mathrsfs}{rsfs}
\DeclareMathAlphabet{\mathcal}{OMS}{cmsy}{m}{n}

\newcommand{\beq}{\begin{equation}}
\newcommand{\eeq}{\end{equation}}

\begin{document}

\title{Les Ondes Gravitationnelles: Une Nouvelle Astronomie\\[0.3cm]
{\small{\rm [Dans ``\textit{O\`u en est-on avec la Relativit\'e\,?}'', \'Eds. Ph. Brax, E. Klein \& P. Vanhove (2018).]}}}

\author{Luc \textsc{Blanchet}}\email{luc.blanchet@iap.fr}
\affiliation{Institut d'Astrophysique de Paris, UMR 7095 du CNRS,
  Sorbonne Universit{\'e}s \& UPMC Universit\'e Paris
  6,\\ 98\textsuperscript{bis} boulevard Arago, 75014 Paris, France}

\date{\today}

\begin{abstract}
L'astronomie contemporaine est en train de vivre une r\'evolution, peut-\^etre plus importante encore que celle qui eu lieu avec l'av\`enement de la radioastronomie dans les ann\'ees 1960, puis l'ouverture du ciel aux observations dans les autres longueurs d'onde \'electromagn\'etiques. Les d\'etecteurs d'ondes gravitationnelles de la collaboration LIGO/Virgo ont observ\'e depuis 2015 les signaux \'emis lors de la collision et la fusion de syst\`emes binaires de trous noirs massifs \`a grande distance astronomique. Cette d\'ecouverte majeure ouvre la voie \`a la nouvelle astronomie des ondes gravitationnelles, radicalement diff\'erente de l'astronomie traditionnelle bas\'ee sur les ondes \'electromagn\'etiques. Plus r\'ecemment, en 2017, la d\'etection d'ondes gravitationnelles \'emises par la fusion d'un syst\`eme binaire d'\'etoiles \`a neutrons a \'et\'e suivie de signaux \'electromagn\'etiques observ\'es par les satellites $\gamma$ et X, et par les observatoires en optique. Une moisson de d\'ecouvertes a \'et\'e possible gr\^ace \`a l'astronomie ``multi-messag\`ere'', qui combine les informations issues de l'onde gravitationnelle avec celles provenant du rayonnement \'electromagn\'etique. Un autre aspect important de la nouvelle astronomie gravitationnelle concerne la physique fondamentale, avec les tests de la relativit\'e g\'en\'erale et des th\'eories alternatives de la gravitation, ainsi que du mod\`ele standard de la cosmologie.
\end{abstract}


\maketitle

\centerline{\textit{Bien \'ecouter, c'est presque r\'epondre}}
\centerline{Marivaux}

\section{\'Ecouter la symphonie de l'Univers}

Depuis les temps imm\'emoriaux --- l'Astronomie est probablement la plus ancienne science humaine --- les hommes ont observ\'e le ciel et \'etudi\'e les astres gr\^ace \`a la lumi\`ere, c'est-\`a-dire, dans le langage moderne, les ondes \'electromagn\'etiques et leur particule associ\'ee, le photon~\cite{Lena96}. Les astronomes analysent cette lumi\`ere avec des instruments toujours plus performants, et l'\'etudient dans toute la gamme des fr\'equences \'electromagn\'etiques, depuis les photons les plus \'energ\'etiques $\gamma$ et X, presque exclusivement observables depuis l'espace, en passant par les infra-rouges, le visible et l'ultra-violet, jusqu'aux ondes radio de grandes longueurs d'onde. Nous n\'egligeons ici les autres porteurs d'information en astronomie que sont les particules de mati\`ere qui forment le rayonnement cosmique, et les neutrinos, qui furent d\'etect\'es par exemple en 1987 lors de la fameuse explosion de supernova dans le grand nuage de Magellan (qui est une petite galaxie satellite de la Voie Lact\'ee). 

\`A chaque fois que les astronomes ont ouvert une ``fen\^etre'' d'observation, en explorant une nouvelle bande de fr\'equences \'electromagn\'etiques, une moisson de d\'ecouvertes eu lieu. Par exemple, la d\'ecouverte des \'emissions radio en provenance du centre de notre Galaxie dans les ann\'ees 1930, puis dans les ann\'ees 1960 la d\'ecouverte de sources tr\`es \'energ\'etiques \'emettant des rayons $\gamma$ et l'explication de la nature des quasars gr\^ace \`a la radioastronomie, puis, plus tard, la d\'ecouverte des pulsars. 

Mais, c'est une \'evidence, si les astronomes ont des yeux (ou plut\^ot, des d\'etecteurs photom\'etriques tr\`es sophistiqu\'es) ils sont ``sourds'', car il n'y a pas de milieu mat\'eriel remplissant l'espace intersid\'eral dans lequel pourraient se propager des ondes sonores. Jusqu'\`a pr\'esent, les astronomes \'etaient comme des explorateurs de la jungle qui observent les arbres de la for\^et et les animaux qui passent \`a proximit\'e, mais ne peuvent pas entendre le bruit du ruissellement de l'eau dans une cascade au loin, ou le rugissement d'un tigre encore plus loin.

La r\'evolution de l'astronomie qui est en train de se produire est que dor\'enavant les astronomes peuvent \'ecouter les objets et les ph\'enom\`enes lointains dans l'Univers gr\^ace aux ondes gravitationnelles\,! En effet celles-ci poss\`edent une analogie profonde avec les ondes sonores. Ainsi, comme les ondes sonores elles sont produites par un mouvement coh\'erent d'ensemble de la mati\`ere ; au contraire, les  ondes \'electromagn\'etiques r\'esultent en g\'en\'eral de la superposition des \'emissions individuelles des atomes et mol\'ecules qui composent la source~\cite{Th300}. La cons\'equence est que la longueur d'onde des ondes gravitationnelles est beaucoup plus grande que la taille de la source qui les \'emet. Cette diff\'erence essentielle avec les ondes \'electromagn\'etiques rend l'astronomie des ondes gravitationnelles tr\`es diff\'erente --- et compl\'ementaire --- de l'astronomie traditionnelle bas\'ee sur la lumi\`ere.

\section{Les vibrations de l'espace-temps} 

Et pourtant, les ondes gravitationnelles se d\'eplacent dans le vide\,! Quel est donc ce milieu qui jouerait un r\^ole analogue \`a celui du milieu mat\'eriel dans lequel se propagent les ondes sonores\,? La r\'eponse, r\'evolutionnaire, fut apport\'ee par la relativit\'e g\'en\'erale d'Einstein de 1915: c'est l'espace-temps lui-m\^eme, qui se courbe au voisinage des masses, et dont les vibrations constituent les ondes gravitationnelles. En relativit\'e g\'en\'erale l'espace-temps n'a pas de caract\`ere absolu et immuable, comme l'espace en avait un en th\'eorie de Newton. C'est une entit\'e dynamique, qui g\'en\'eralise l'espace-temps ``plat'' de Minkowski, \`a la base de la relativit\'e restreinte formul\'ee en 1905 par Einstein, Lorentz et Poincar\'e. En relativit\'e g\'en\'erale, la sensation famili\`ere du champ gravitationnel est interpr\'et\'ee comme l'effet de la d\'eformation de l'espace-temps, courb\'e au sens de la courbure riemannienne invent\'ee par Riemann en 1854. Cette conception du r\^ole de l'espace-temps vient des fondements m\^emes de la relativit\'e g\'en\'erale ; c'est la cons\'equence du principe d'\'equivalence, traduction moderne du fait que tous les corps, ind\'ependamment de leur structure et composition interne, sont acc\'el\'er\'es de la m\^eme fa\c{c}on dans un champ de gravitation~\cite{Will}.

La relativit\'e g\'en\'erale englobant les principes de la relativit\'e restreinte, les vibrations de l'espace-temps ou ondes gravitationnelles se propagent \`a la vitesse de la lumi\`ere. De plus elles suivent les m\^emes trajectoires que les ondes \'electromagn\'etiques dans l'espace-temps ; elles sont d\'evi\'ees, comme la lumi\`ere, par les champs gravitationnels, produits par exemple par les amas de galaxies et les grandes structures dans l'Univers. Par contre, elles interagissent extr\^emement faiblement avec la mati\`ere, et peuvent donc se propager sans impunit\'e sur des tr\`es grandes distances. Notons \`a ce propos qu'il est impossible d'``enfermer'' une onde gravitationnelle dans une boite, comme on pourrait le faire avec une onde \'electromagn\'etique, qui peut subir des r\'eflexions multiples dans une boite garnie int\'erieurement de miroirs. En effet il n'existe aucun \'ecran \`a une onde gravitationnelle, car les masses, qui engendrent les champs gravitationnels, sont toujours positives en relativit\'e g\'en\'erale: on ne peut pas annuler le champ gravitationnel. On s'attend donc \`a ce que les ondes gravitationnelles nous livrent un tr\'esor d'informations en astronomie, \`a des tr\`es grandes distances et des \'epoques recul\'ees dans le pass\'e, probablement jusque juste apr\`es le Big-Bang. 

\section{La relativit\'e g\'en\'erale \'etait correcte\,!}

La relativit\'e g\'en\'erale est une th\'eorie c\'el\`ebre non seulement pour l'\'el\'egance de ses \'equations, mais pour son caract\`ere incroyablement pr\'edictif. Toutes les pr\'edictions de cette th\'eorie se sont av\'er\'ees correctes, c'est-\`a-dire v\'erifi\'ees par des observations ou bien exp\'erimentalement: d\'eviation de la lumi\`ere par le Soleil, lentilles et arcs gravitationnels en astronomie, le retard gravitationnel de la lumi\`ere ou effet Shapiro (mesur\'e en 1964), le d\'ecalage gravitationnel des fr\'equences ou effet Einstein qui joue un r\^ole important pour la pr\'ecision du syst\`eme GPS~\cite{Will}. Seul l'effet de pr\'ecession anormale du mouvement de la plan\`ete Mercure, qui avait \'et\'e mis en \'evidence par Le Verrier en 1859, a plut\^ot \'et\'e une ``post-diction'' de la th\'eorie. Quant aux ondes gravitationnelles, leur existence a d'abord \'et\'e \'etablie par des travaux th\'eoriques en relativit\'e g\'en\'erale avant d'\^etre confirm\'ee par les observations\,!

Mais \'etablir la r\'ealit\'e physique des ondes gravitationnelles ne fut pas une mince affaire. Dans son article de 1916~\cite{E16}, Einstein montra que la m\'etrique de l'espace-temps ob\'eit \`a une \'equation des ondes dans un syst\`eme de coordonn\'ees particulier. Mais quelle est la signification des ondes gravitationnelles si leur description d\'epend d'un syst\`eme de coordonn\'ees, qu'un observateur est libre de choisir de fa\c{c}on arbitraire\,? De nos jours on r\'esout le probl\`eme en montrant que la courbure de l'espace-temps se propage et a bien, elle, une signification physique intrins\`eque, ind\'ependante d'un observateur. Une autre question a \'et\'e de savoir si les ondes gravitationnelles transportent de l'\'energie, qui serait extraite de sa source et peut \^etre d\'epos\'ee sur un d\'etecteur\,? Enfin, les effets de r\'eaction \`a l'\'emission des ondes gravitationnelles qui s'exercent dans la source (par exemple le recul de la source r\'esultant d'une \'emission anisotrope d'ondes gravitationnelles) sont-ils coh\'erents avec les flux d'ondes mesur\'es \`a grandes distances de la source\,? La r\'eponse \`a toutes ces questions est ``oui'', comme \'etabli dans les ann\'ees 1960 par de nombreux travaux th\'eoriques et des exp\'eriences de pens\'ee~\cite{Bondi57,BBM62,Ehletal76}.

\begin{figure}[h]
\begin{center}
\includegraphics[width=10cm]{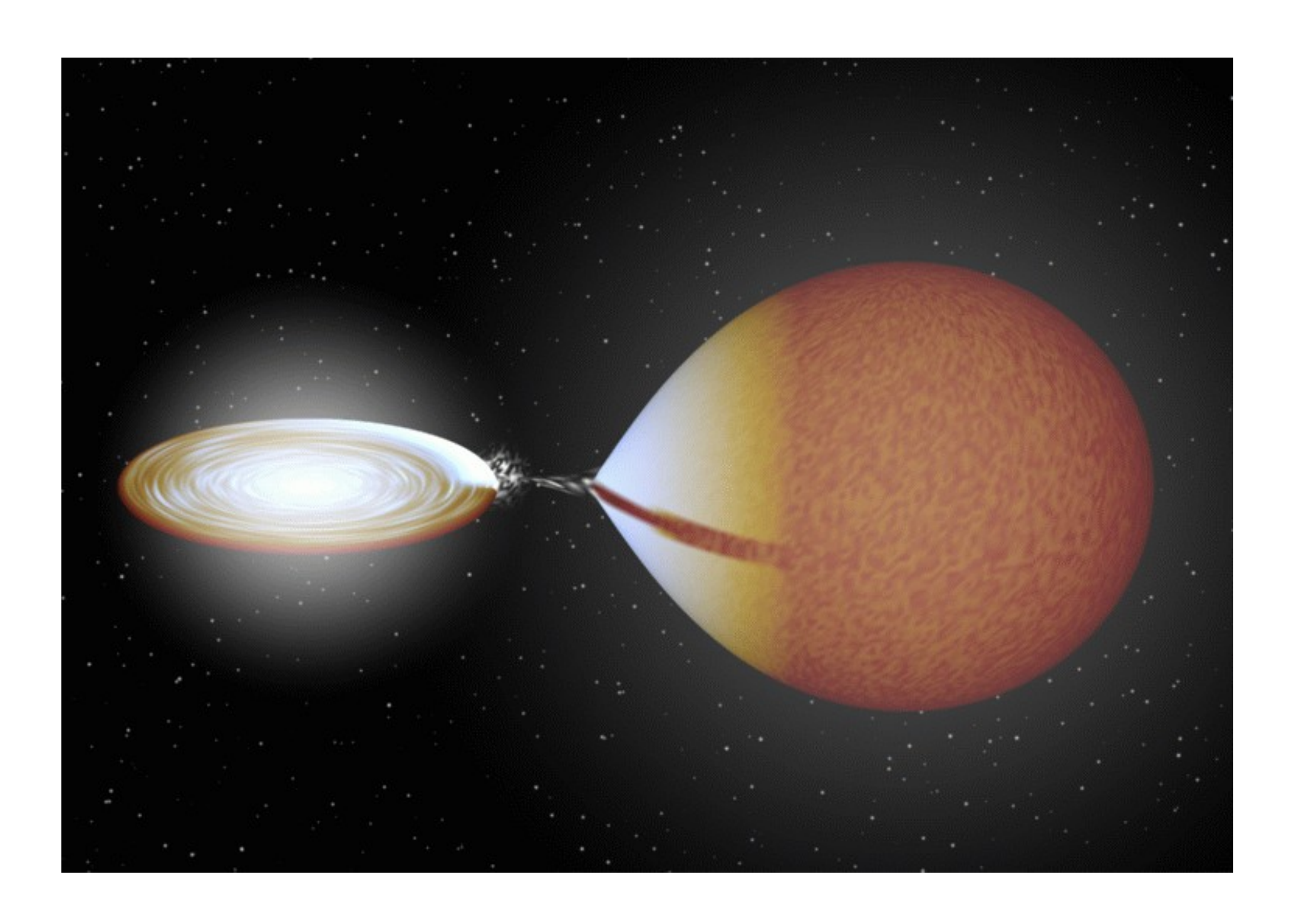}
\end{center}
\caption{Une binaire ``cataclysmique''. L'\'etoile en fin d'\'evolution stellaire d\'everse de la mati\`ere sur un disque d'accr\'etion entourant une naine blanche. Les \'emissions X et UV venant du disque d'accr\'etion permettent d'\'etudier la dynamique de ces syst\`emes et notamment de mesurer le transfert de masse entre les deux \'etoiles.}\label{fig1}
\end{figure}

\begin{figure}[h]
\begin{center}
\includegraphics[width=10cm]{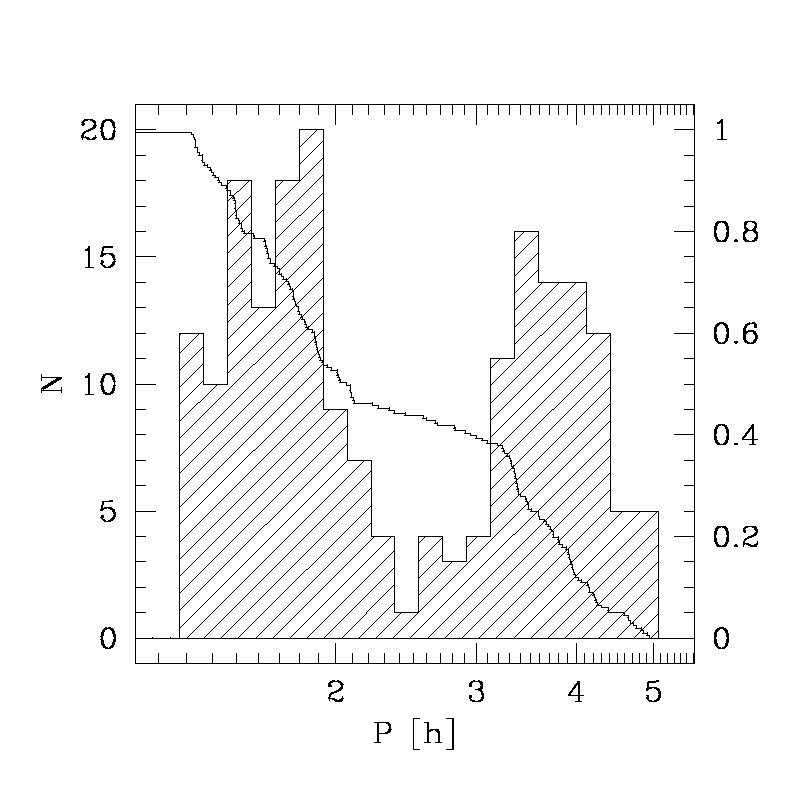}
\end{center}
\caption{Histogramme des binaires cataclysmiques en fonction de la p\'eriode orbitale. La pr\'esence du pic pour des p\'eriodes orbitales inf\'erieures \`a 2 heures s'explique par un m\'ecanisme dans lequel le rayonnement gravitationnel \'emis par le syst\`eme joue un r\^ole.}\label{fig2}
\end{figure}

La gravitation \'etant de loin la plus faible des quatre forces fondamentales connues, aucune source terrestre d'ondes gravitationnelles n'est assez puissante pour g\'en\'erer un signal observable. Il faut se tourner vers des sources cosmiques: d\`es les ann\'ees 1960 une classe de syst\`emes binaires d'\'etoiles dans notre Galaxie avaient d\'ej\`a donn\'e certaines indications sur l'existence du rayonnement gravitationnel. Dans ces syst\`emes, dits ``cataclysmiques'', l'une des \'etoiles, qui est en fin d'\'evolution stellaire, remplit son lobe de Roche et d\'everse de la mati\`ere pour former un disque d'accr\'etion autour de son compagnon, qui est une \'etoile compacte plus massive appel\'ee naine blanche (voir la figure~\ref{fig1}). On observe la binaire par l'\'emission UV et X produite par la mati\`ere chauff\'ee par le fort champ gravitationnel de la naine blanche. Dans certains cas, pour des p\'eriodes orbitales tr\`es courtes, ces syst\`emes binaires ne sont stables que si l'on invoque une perte de moment cin\'etique orbital par \'emission d'ondes gravitationnelles (voir la figure~\ref{fig2})~\cite{Kraft62,Pacz67,Faulk71}. Ainsi, les astronomes ont tr\`es t\^ot utilis\'e le rayonnement gravitationnel pour expliquer l'existence de certains syst\`emes astrophysiques\,!


\begin{figure}[h]
\begin{center}
\includegraphics[width=8cm]{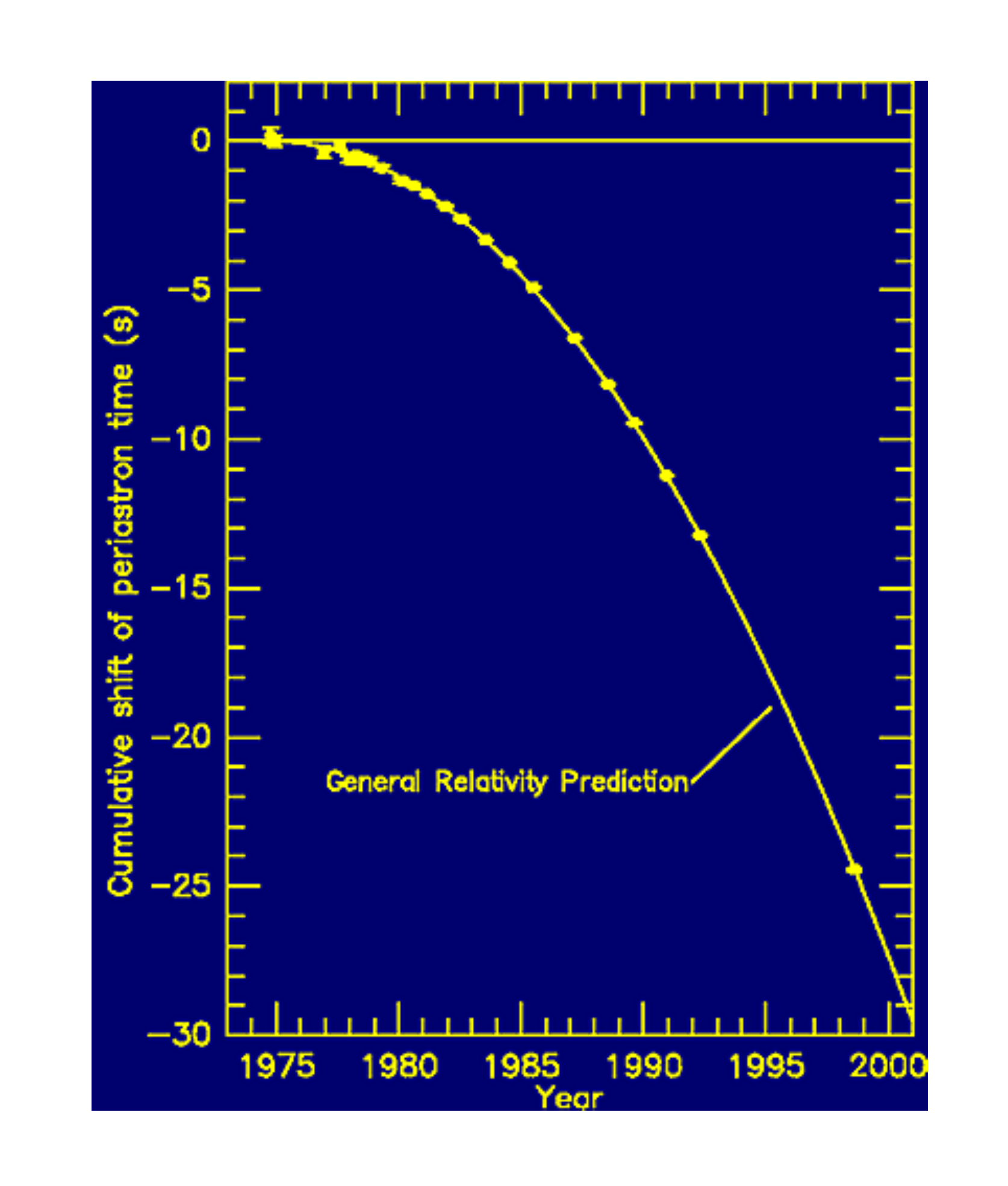}
\end{center}
\caption{La d\'ecroissance de la phase orbitale du pulsar binaire, en parfait accord avec la pr\'ediction de la relativit\'e g\'en\'erale et de la formule du quadrup\^ole d'Einstein pour l'\'emission de rayonnement gravitationnel~\cite{TW82}.}\label{fig3}
\end{figure}

Mais c'est bien s\^ur gr\^ace au pulsar binaire, d\'ecouvert par Hulse et Taylor en 1974~\cite{HulseTaylor}, que la premi\`ere mise en \'evidence pr\'ecise des ondes gravitationnelles eu lieu. Le pulsar, qui est une \'etoile \`a neutrons fortement magn\'etis\'ee en rotation rapide sur elle-m\^eme (ici avec une p\'eriode de 56 ms), envoie \`a chaque rotation un pulse de rayonnement \'electromagn\'etique radio en direction de la Terre. L'analyse des instants d'arriv\'ee des pulses montre, gr\^ace \`a l'effet Doppler, que le pulsar est en orbite autour d'une autre \'etoile, probablement aussi une \'etoile \`a neutrons. En 1979, apr\`es plusieurs ann\'ees d'observations, \'etait d\'etect\'ee une diminution s\'eculaire de la p\'eriode orbitale du syst\`eme binaire --- variation qui s'explique parfaitement avec la relativit\'e g\'en\'erale, par la perte d'\'energie li\'ee \`a l'\'emission du rayonnement gravitationnel, voir la figure~\ref{fig3}.

En fait la pr\'ediction de la relativit\'e g\'en\'erale n'est pas ``exacte'', car il est impossible d'obtenir une solution exacte des \'equations d'Einstein (qui sont tr\`es ``non-lin\'eaires'') pour le probl\`eme \`a deux corps~\cite{MTW}. On utilise donc une m\'ethode d'approximation, qui consiste \`a d\'evelopper la th\'eorie selon une s\'erie en puissances d'un petit param\`etre, qui est le rapport entre la vitesse des corps sur leur orbite et la vitesse de la lumi\`ere. C'est ce qu'on appelle le d\'eveloppement ``post-newtonien''. Le rapport $v/c$ est r\'eellement petit dans le cas du pulsar binaire, de l'ordre de un milli\`eme, donc l'approximation post-newtonienne va \^etre tr\`es bonne. \`A l'ordre dominant, l'onde gravitationnelle a une forme quadrupolaire (alors que l'onde \'electromagn\'etique est en g\'en\'eral dipolaire), et est d\'ecrite par la formule du quadrup\^ole d'Einstein de 1918~\cite{E18}. Celle-ci donne entre autres l'\'energie totale \'emise par le syst\`eme binaire sous forme d'ondes gravitationnelles, et est donc tr\`es bien test\'ee par la d\'ecroissance orbitale du pulsar binaire.

\section{Les syst\`emes binaires compacts spiralants}

Lorsqu'\`a la fin de sa vie, dans environ 350 millions d'ann\'ees, le pulsar binaire et son compagnon auront \'emis toute leur \'energie gravitationnelle, ils deviendront ce qu'on appelle une binaire compacte ``spiralante''. Dans un tel syst\`eme, form\'e soit d'\'etoiles \`a neutrons soit de trous noirs, les deux corps compacts suivent une orbite rapproch\'ee, dans les derniers instants pr\'ec\'edant leur fusion finale, qui est une spirale rentrante \`a cause de la perte d'\'energie li\'ee \`a l'\'emission des ondes gravitationnelles. La fr\'equence orbitale et l'amplitude du signal augmentent au cours du temps, jusqu'\`a ce que le syst\`eme atteigne la derni\`ere orbite circulaire stable, apr\`es laquelle les deux corps entrent en collision et fusionnent rapidement pour former un trou noir. Celui-ci est initialement d\'eform\'e et hautement dynamique \`a cause de la collision, mais il finit par atteindre, par \'emission en ondes gravitationnelles de ses modes dynamiques, un r\'egime stationnaire d\'ecrit par la solution du trou noir de Kerr. Cette solution des \'equations d'Einstein, d\'ecouverte par Kerr en 1963~\cite{Kerr63}, repr\'esente le trou noir le plus g\'en\'eral non charg\'e et en rotation --- ici la rotation provient du mouvement orbital des deux corps avant la collision, et de la loi de conservation du moment cin\'etique.

\begin{figure}[h]
\begin{center}
\includegraphics[width=10cm]{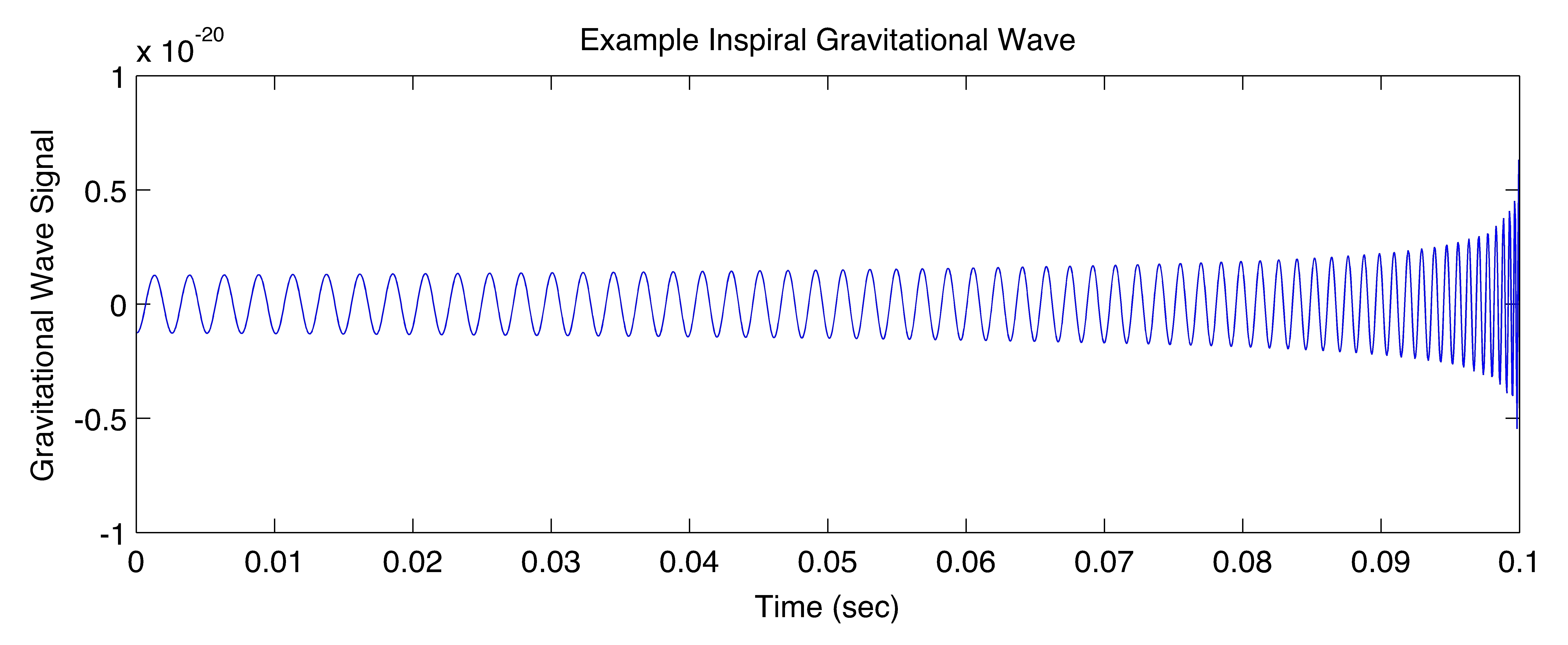}
\end{center}
\caption{Le signal gravitationnel produit par un syst\`eme binaire d'\'etoiles compactes dans la phase spiralante pr\'ec\'edant la coalescence finale. Pendant cette phase le signal est ``universel'': il ne d\'epend que de la masse des corps mais pas de leur nature, que ce soient des \'etoiles \`a neutrons, des trous noirs ou peut-\^etre, des objets plus ``exotiques'' tels que les \'etoiles bosoniques~\cite{RuffBona69}.}\label{fig4}
\end{figure}

L'observation de l'onde gravitationnelle dans la phase spiralante permet de mesurer les masses s\'epar\'ees des deux objets compacts et leur rotation intrins\`eque ou spin. En particulier, une certaine combinaison des deux masses, appel\'ee masse de ``gazouillement'' (ou masse ``chirp''), est tr\`es bien d\'etermin\'ee car elle intervient dans la formule du quadrup\^ole, c'est-\`a-dire \`a l'ordre dominant. La figure~\ref{fig4} montre le signal gravitationnel produit lors de la phase spiralante de deux objets compacts.

\begin{figure}[h]
\begin{center}
\includegraphics[width=12cm]{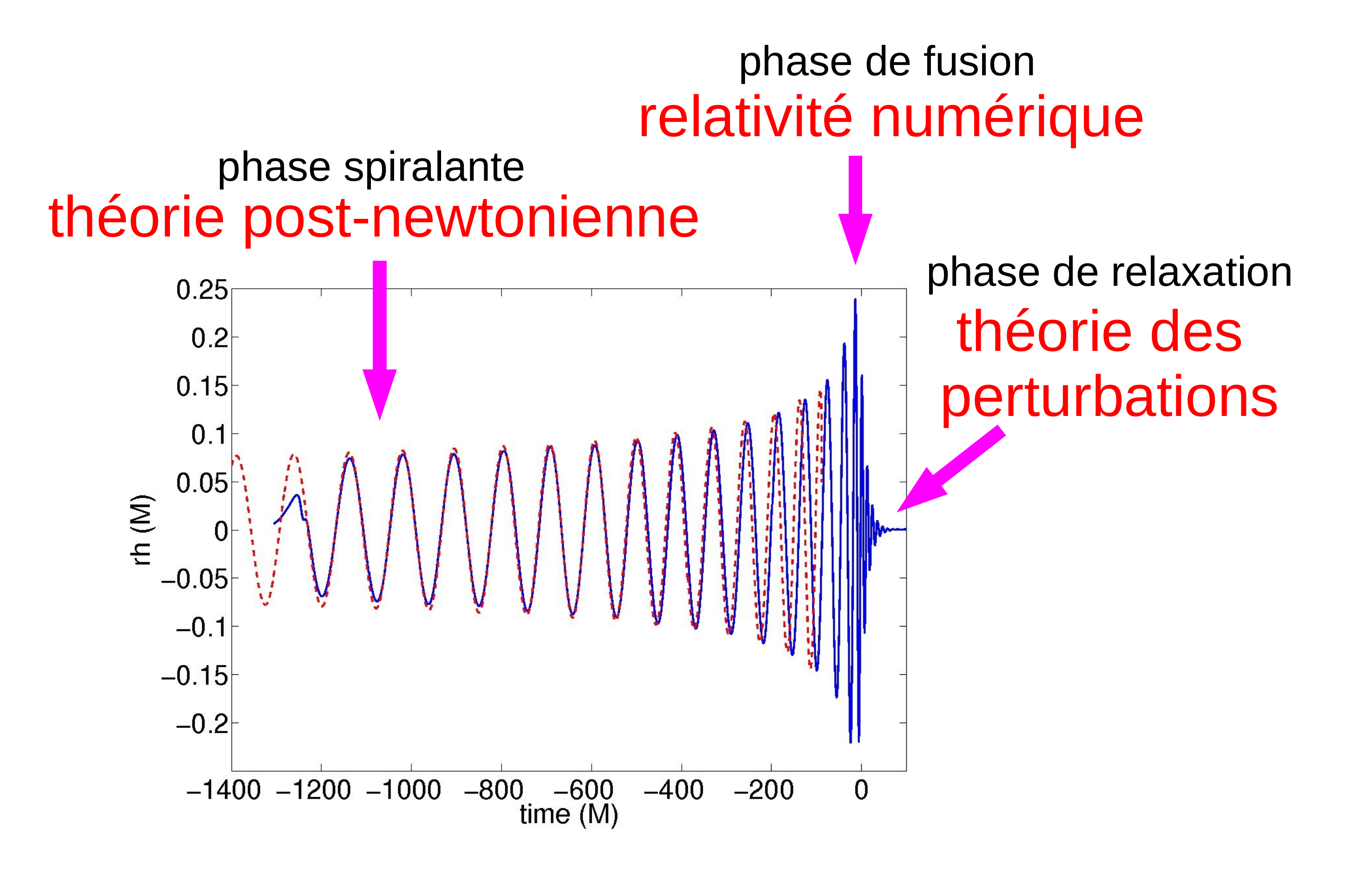}
\end{center}
\caption{L'onde gravitationnelle dans les trois phases de la coalescence d'un syst\`eme de deux trous noirs. Dans la phase spiralante le signal est calcul\'e par des m\'ethodes post-newtoniennes en relativit\'e g\'en\'erale (pointill\'es rouges). Le calcul post-newtonien se raccorde avec le calcul num\'erique valable dans les phases de fusion et de relaxation (trait bleu).}\label{fig5}
\end{figure}

Pendant la phase spiralante, l'onde gravitationnelle est calcul\'ee gr\^ace \`a l'approximation post-newtonienne de la relativit\'e g\'en\'erale (d\'eveloppement quand $v/c$ est petit)~\cite{BlanchetLR}. Dans le cas de deux \'etoiles \`a neutrons, cette approximation est excellente, mais elle doit avoir \'et\'e d\'evelopp\'ee \`a un ordre \'elev\'e, pour pouvoir suivre avec grande pr\'ecision les milliers de cycles du signal dans la bande de fr\'equence des d\'etecteurs. Au moment de la fusion l'approximation post-newtonienne n'est plus valable et est remplac\'ee par un calcul num\'erique des \'equations d'Einstein. Le probl\`eme du calcul num\'erique a repr\'esent\'e un d\'efi pendant de longues ann\'ees mais a maintenant \'et\'e r\'esolu~\cite{Pret05,Bak06,Camp06}. D'autre part, des m\'ethodes effectives telles que ``effective-one-body'' (EOB) permettent d'interpoler de fa\c{c}on analytique entre la phase post-newtonienne et le calcul num\'erique~\cite{BuonD99}. Finalement la phase finale, pendant laquelle le trou noir form\'e se ``relaxe'' vers la solution de Kerr, est analys\'ee par une m\'ethode de perturbation de trou noir. La figure~\ref{fig5} montre les phases successives de la coalescence, ainsi que les m\'ethodes th\'eoriques utilis\'ees pour les d\'ecrire.

\section{La d\'etection des ondes gravitationnelles}

Le premier d\'etecteur d'ondes gravitationnelles fut con\c{c}u par Weber dans les ann\'ees 1960. Il s'agissait d'un cylindre m\'etallique r\'esonant, sur laquelle l'onde gravitationnelle d\'epose de l'\'energie, et dont les oscillations m\'ecaniques sont converties en signal \'electrique par un transducteur. Des ``barres de Weber'' ont \'et\'e en service jusque dans les ann\'ees 2000 mais ont \'et\'e abandonn\'ees \`a cause de leur sensibilit\'e trop faible et, surtout, de leur bande de fr\'equence \'etroite, limit\'ee essentiellement \`a la fr\'equence de r\'esonance de la barre. 

\begin{figure}[h]
\begin{center}
\includegraphics[width=12cm]{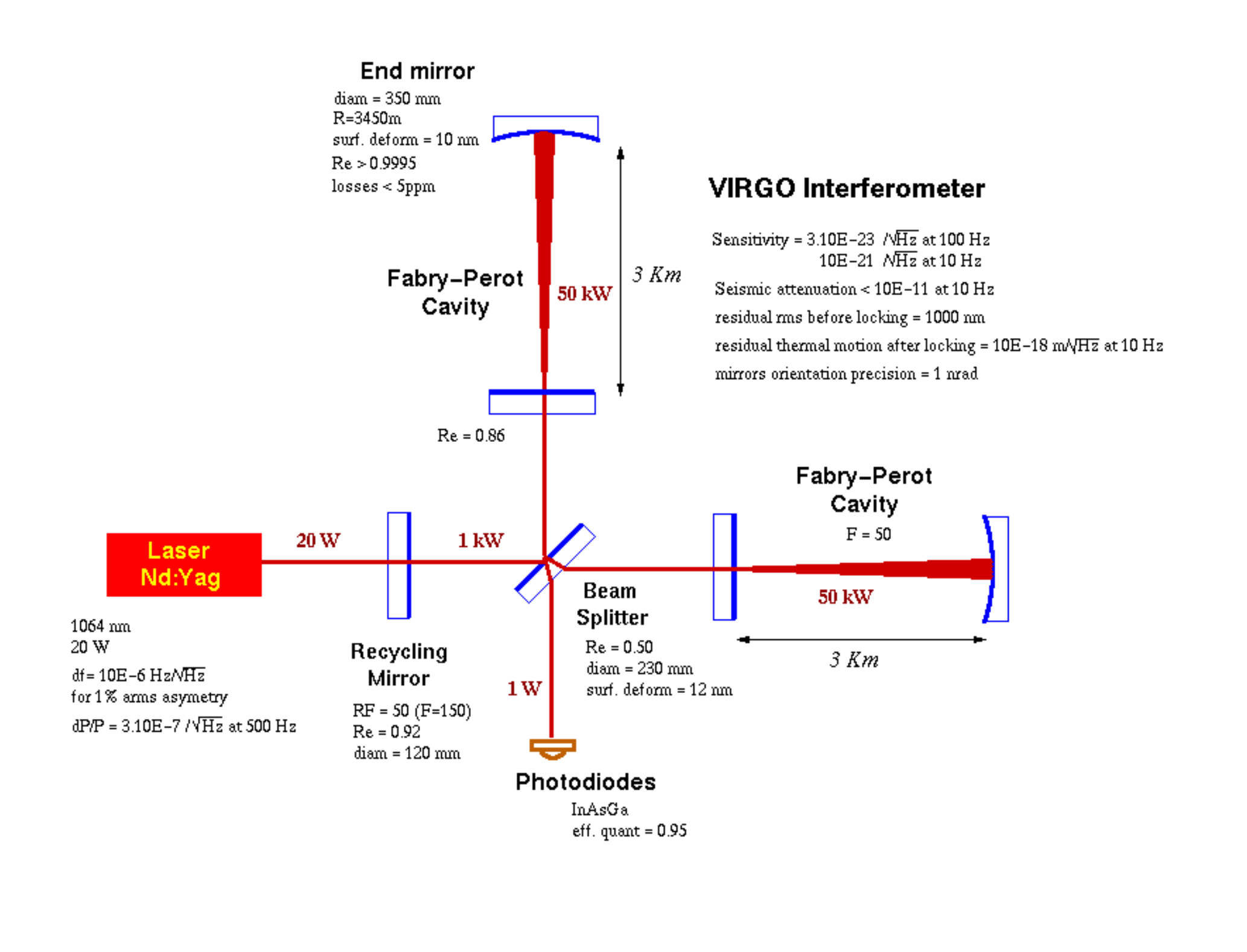}
\end{center}
\caption{Principe de l'interf\'erom\`etre \`a laser pour la d\'etection du rayonnement gravitationnel. Les deux bras sont des cavit\'es r\'esonantes Fabry-Perot qui emmagasinent la puissance \'electromagn\'etique. La s\'eparatrice et les miroirs sont suspendus \`a un isolateur sismique. Un miroir \`a l'entr\'ee permet de recycler la lumi\`ere vers l'interf\'erom\`etre.}\label{fig6}
\end{figure}

Les d\'etecteurs actuels sont des interf\'erom\`etres \`a laser de type Michelson-Morley, dont les bras ont plusieurs kilom\`etres de long, et sont constitu\'es de cavit\'es laser r\'esonantes de Fabry-Perot. La lame s\'eparatrice, qui s\'epare le faisceau laser incident en deux faisceaux dirig\'es vers les deux bras, et les miroirs d'entr\'ee/sortie et en bout de cavit\'es, sont suspendus \`a un syst\`eme pendulaire qui permet de s'affranchir des vibrations sismiques terrestres aux fr\'equences basses. L'onde gravitationnelle est d\'etect\'ee par la variation de la diff\'erence de chemin optique entre les deux bras. Outre le bruit sismique, les principales sources de bruit dans l'interf\'erom\`etre sont le bruit thermique des miroirs et le ``bruit de photon'' des lasers. Le grand int\'er\^et de l'interf\'erom\`etre laser est sa large bande de fr\'equence, depuis environ 30 Hz jusqu'\`a quelques milliers de Hz. Voir la figure~\ref{fig6}.

\begin{figure}[h]
\begin{center}
\includegraphics[width=10cm]{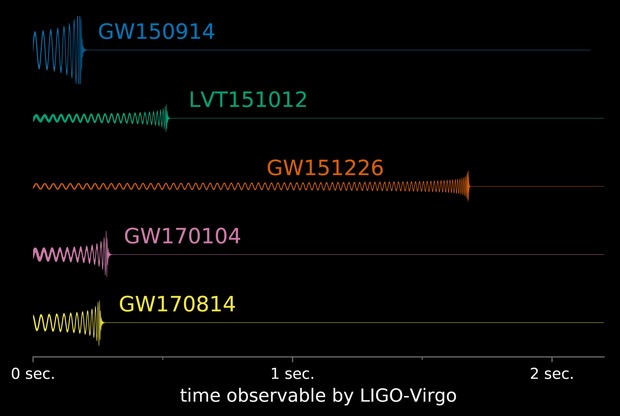}
\end{center}
\caption{Les ondes gravitationnelles issues de coalescences de trous noirs observ\'ees jusqu'\`a pr\'esent par la collaboration LIGO-Virgo. Les \'ev\'enements sont d\'esign\'es par ``Gravitational Wave'' (GW) suivi de la date. Ainsi, le 1er \'ev\'enement, du 14 septembre 2015, est appel\'e GW150914. Le second \'ev\'enement, du 12 octobre 2015, \'etait beaucoup plus incertain, car, s'il \'etait r\'eel, beaucoup plus loin, et est appel\'e ``LIGO-Virgo Trigger'' (LVT). L'\'ev\'enement GW170814 a \'et\'e vu par les trois d\'etecteurs du r\'eseau LIGO-Virgo.}\label{fig7}
\end{figure}

Le 14 septembre 2015 les d\'etecteurs d'ondes gravitationnelles LIGO ont observ\'e le signal de la coalescence de deux trous noirs stellaires de masses 36 et 29 fois la masse du Soleil, \`a une distance de environ 400 Mpc~\cite{GW150914}. L'\'ev\'enement a dur\'e une fraction de seconde, durant laquelle les deux trous noirs ont parcouru les quelques derniers cycles orbitaux \`a une vitesse proche de la lumi\`ere avant de fusionner. Le signal a \'et\'e d\'etect\'e avec un fort signal-\`a-bruit de 23 et un taux de fausse alarme \'equivalent \`a un niveau de confiance de 5,1 s. Le signal est arriv\'e dans les deux d\'etecteurs LIGO, situ\'es sur les c\^otes Ouest et Est am\'ericaines, avec un \'ecart de 6,5 millisecondes. Cette diff\'erence des temps d'arriv\'ee, compatible avec la distance lumi\`ere entre les deux d\'etecteurs (10 millisecondes), fournit une indication sur la direction de l'\'ev\'enement, qui a \'et\'e localis\'e dans une portion d'arc de cercle dans le ciel d'environ 600 degr\'es carr\'es. A l'\'epoque, le d\'etecteur europ\'een Virgo n'\'etait malheureusement pas en fonctionnement, et la pr\'ecision sur la localisation de la source avec seulement deux d\'etecteurs n'\'etait pas tr\`es bonne. Depuis, plusieurs autres \'ev\'enements de coalescences de trous noirs ont \'et\'e observ\'es, voir la figure~\ref{fig7}. Le 14 ao\^ut 2017 un \'ev\'enement de trous noirs \'etait observ\'e mais cette fois avec les trois d\'etecteurs LIGO et Virgo simultan\'ement. La boite d'erreur sur la localisation dans le ciel a donc \'et\'e drastiquement r\'eduite, \`a environ 30 degr\'es carr\'es. L'astronomie gravitationnelle \'etait n\'ee\,!

Une surprise est que les masses mesur\'ees des trous noirs sont tr\`es \'elev\'ees, en g\'en\'eral bien sup\'erieures \`a celles auxquelles les astronomes \'etaient habitu\'es, pour les trous noirs connus dans notre galaxie par l'\'emission X du disque d'accr\'etion de mati\`ere provenant d'une \'etoile compagnon. Expliquer la formation de trous noirs aussi massifs par l'explosion d'une \'etoile en fin d'\'evolution stellaire, puis trouver un m\'ecanisme qui induit une binaire de deux de ces trous noirs assez serr\'ee pour \'evoluer par rayonnement gravitationnel dans moins d'un temps de Hubble, repr\'esentent des d\'efis pour l'astrophysique non encore r\'esolus actuellement. 

Lors de la fusion des deux trous noirs, une fraction de la masse-\'energie du syst\`eme binaire, prise sur l'\'energie de liaison gravitationnelle, est transf\'er\'ee \`a l'onde gravitationnelle. Lors du premier \'ev\'enement (appel\'e ``GW150914''), trois masses solaires ont ainsi \'et\'e \'emises sous forme d'onde gravitationnelle en une fraction de seconde, ce qui en fait l'\'ev\'enement le plus puissant jamais observ\'e, avec une puissance d'environ $10^{49}$ watts\,! Par contre, les syst\`emes de trous noirs n'\'emettent aucune autre forme d'\'energie, car le  trou noir est une solution purement gravitationnelle des \'equations d'Einstein. Dans l'astronomie gravitationnelle on peut entendre les collisions de trous noirs mais on ne les voit pas\,!

\section{L'astronomie multi-messagers}

\begin{figure}[h]
\begin{center}
\includegraphics[width=12cm]{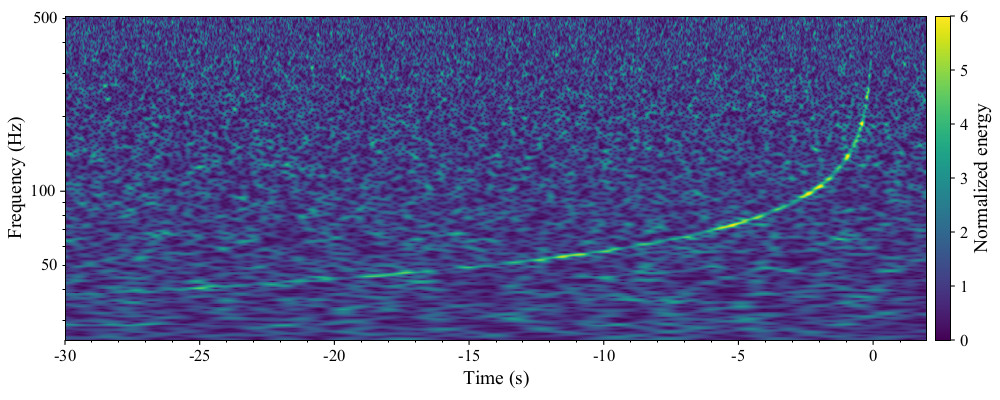}
\end{center}
\caption{L'augmentation en fr\'equence du train d'ondes gravitationnelles avec le temps, produite alors que les deux \'etoiles \`a neutrons se rapprochent en spiralant l'une autour de l'autre. Le signal se termine par la fusion proprement dite, qui intervient \`a haute fr\'equence, en dehors de la bande de sensibilit\'e des d\'etecteurs actuels.}\label{fig8}
\end{figure}

Le 17 ao\^ut 2017, survenait un autre type d'\'ev\'enement, peut-\^etre encore plus fantastique: la coalescence d'un syst\`eme binaire de deux \'etoiles \`a neutrons~\cite{GW170817}. En effet, les masses mesur\'ees sont beaucoup plus faibles, la masse ``chirp'' \'etant d'environ 1,2 masse solaire, coh\'erente avec les masses attendues pour des \'etoiles \`a neutrons. Malgr\'e les masses plus faibles, le signal-\`a-bruit, de l'ordre de 32, est encore plus fort que les d\'etections de trous noirs, car la distance de l'\'ev\'enement est beaucoup plus proche: 40 Mpc. La figure~\ref{fig8} donne le signal observ\'e pour GW170817 dans le plan ``temps-fr\'equence'', montrant la mont\'ee en fr\'equence caract\'eristique du chirp gravitationnel.

Dans les derni\`eres rotations orbitales, les deux \'etoiles \`a neutrons sont d\'eform\'ees \`a cause des forces de mar\'ees. L'observation de l'onde gravitationnelle a d\'ej\`a permis de placer une contrainte sur le taux de d\'eformation ce qui est tr\`es int\'eressant car l'effet d\'epend des d\'etails de la structure interne des \'etoiles \`a neutrons, en particulier de l'\'equation d'\'etat qui r\`egne au coeur des \'etoiles \`a neutrons. On obtient donc pour la premi\`ere fois une information sur l'\'etat de la mati\`ere nucl\'eaire dans les \'etoiles \`a neutrons. 

\begin{figure}[h]
\begin{center}
\includegraphics[width=12cm]{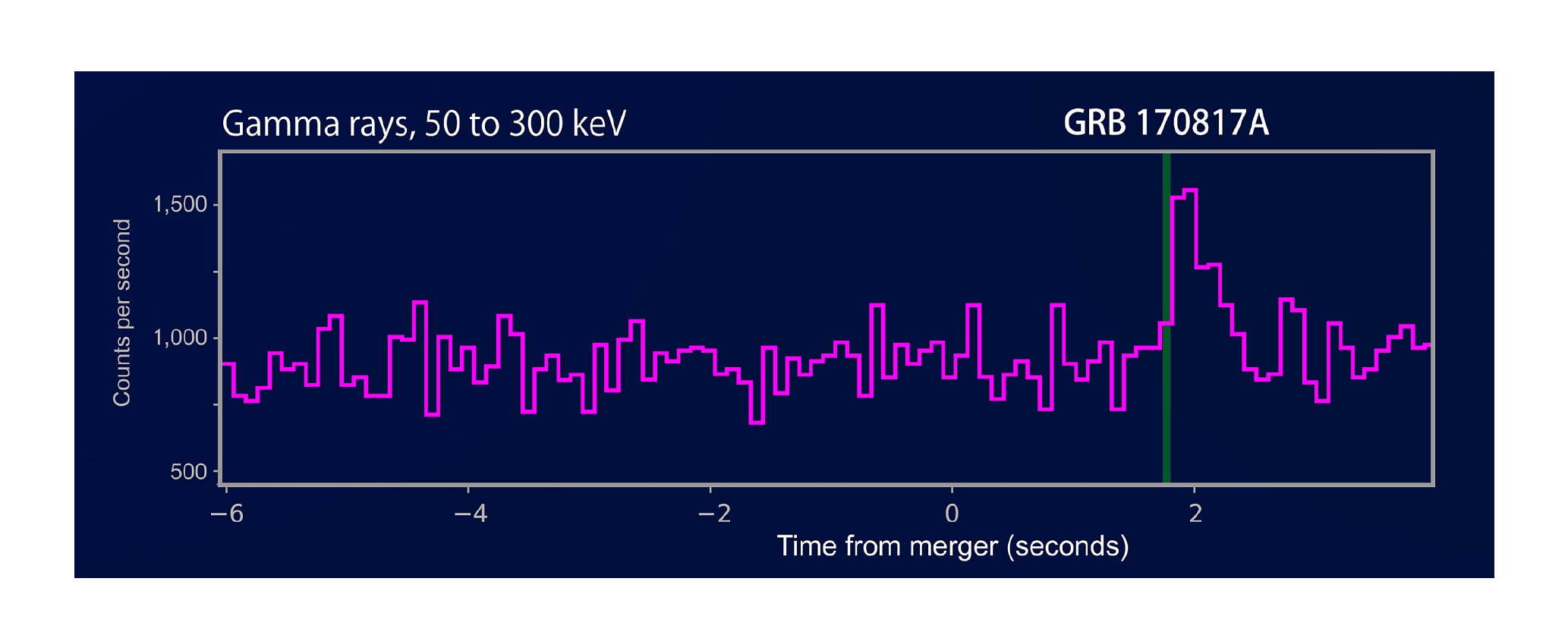}
\end{center}
\caption{Le sursaut $\gamma$ vu par le satellite Fermi 1,7 seconde apr\`es la fin du signal gravitationnel. Le sursaut $\gamma$ est tr\`es \'energ\'etique et provient d'un jet de mati\`ere relativiste tr\`es focalis\'e. Dans le cas pr\'esent il a une relativement faible luminosit\'e ce qui sugg\`ere qu'il a \'et\'e vu ``de c\^ot\'e'' par rapport \`a l'axe du jet relativiste (qui est perpendiculaire au plan orbital des deux \'etoiles \`a neutrons).}\label{fig9}
\end{figure}

Environ 1,7 seconde apr\`es la fusion des deux \'etoiles \`a neutrons, \'etait d\'etect\'e un sursaut \'electromagn\'etique $\gamma$ par les satellites Fermi et Integral. L'astronomie multi-messagers \'etait n\'ee~\cite{LIGOmultmess}\,! Voir la figure~\ref{fig9}.

Cette diff\'erence de 1,7s entre les temps d'arriv\'ee de l'onde gravitationnelle et du sursaut $\gamma$, est hautement significative. Elle est due tr\`es probablement \`a la diff\'erence des instants d'\'emission des deux types d'ondes au niveau de la source. Elle s'interpr\`ete dans le cadre d'un mod\`ele de sursaut $\gamma$, comme le temps mis par l'\'energie \'electromagn\'etique pour s'\'echapper, \`a partir d'un jet de mati\`ere relativiste, \'emis perpendiculairement au plan orbital de la collision. Dans un mod\`ele particulier dit du ``cocon'', le jet relativiste lui-m\^eme doit percer, tel une foreuse, les couches de mati\`ere dense issues des d\'ebris de l'explosion, avant d'engendrer le sursaut $\gamma$. De nombreuses \'etudes sont en cours pour pr\'eciser les param\`etres et la g\'eom\'etrie de l'explosion. Dans l'\'ev\'enement GW170817 le sursaut $\gamma$ a probablement \'et\'e vu sous un angle d'environ 30 degr\'es par rapport \`a l'axe du jet relativiste.

\begin{figure}[h]
\begin{center}
\includegraphics[width=12cm]{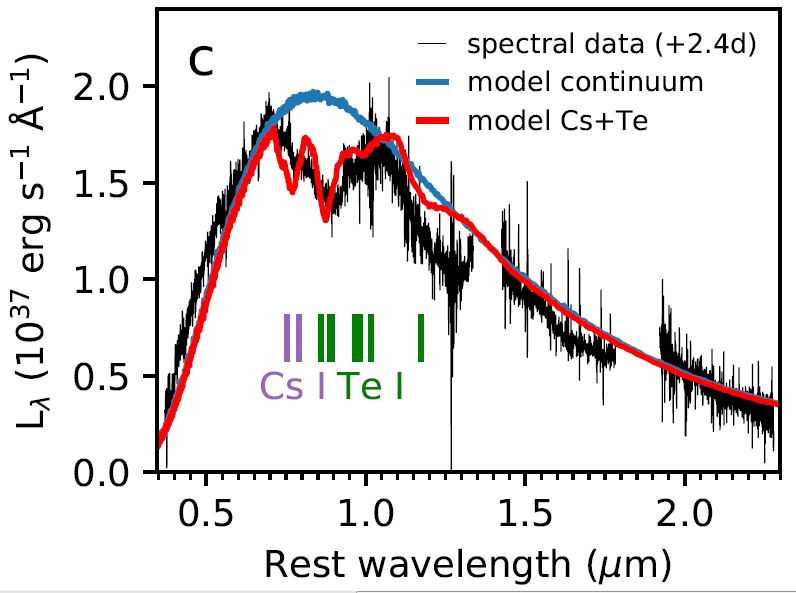}
\end{center}
\caption{Le spectre de la kilonova. C'est essentiellement un spectre de corps noir, avec une temp\'erature de l'ordre de 6000 degr\'es, mais avec des raies d'absorption dues au Tellure et au C\'esium, deux \'el\'ements lourds (num\'eros atomique 52 et 55) dont la formation est difficile \`a expliquer dans les explosions de supernovas. Les raies sont \'elargies par effet Doppler d\^u \`a la vitesse d'\'ejection d'environ 60\,000 km/s.}\label{fig10}
\end{figure}

Peut-\^etre la d\'ecouverte la plus importante fut l'identification, gr\^ace \`a un grand nombre de t\'elescopes observant notamment en optique, et mobilisant une grande partie de la communaut\'e astronomique, de la source des ondes gravitationnelles et du sursaut $\gamma$. En effet, dans les heures et jours qui suivirent l'\'ev\'enement, \'etait identifi\'e dans la boite d'erreur indiqu\'ee par les d\'etecteurs gravitationnels, un objet transitoire en association avec la galaxie NGC 4993. Cet objet, vu en optique, infrarouge, radio et rayons X, n'\'etait pas pr\'esent lors de campagnes d'observations pr\'ec\'edentes. La galaxie NGC 4993 est bien situ\'ee \`a une distance d'environ 40 Mpc, en accord avec l'estimation des d\'etecteurs gravitationnels. L'objet transitoire a rapidement \'et\'e identifi\'e \`a ce qu'on appelle une ``kilonova''.

\`A l'origine, la kilonova \'etait un mod\`ele  th\'eorique d\'ecrivant l'explosion r\'esultant de la  coalescence de deux \'etoiles \`a neutrons~\cite{Metzger10}. Bien que l'\'etoile \`a neutrons soit constitu\'ee essentiellement de neutrons, sa cro\^ute contient des empilements de noyaux lourds comme le fer ou le nickel, se r\'epartissant selon leur densit\'e en une structure cristalline. Certains de ces noyaux lourds ont \'et\'e synth\'etis\'es lors de l'explosion en supernova qui a donn\'e lieu \`a la formation de l'\'etoile \`a neutrons. Lors de la fusion des deux \'etoiles la majeure partie de la masse s'effondre pour former le trou noir. Mais une petite fraction des neutrons sont \'eject\'es et se combinent alors avec les noyaux lourds existants pour former des noyaux beaucoup plus lourds, tr\`es riches en neutrons et instables. Ceux-ci se d\'esint\`egrent alors rapidement, et peuvent former des noyaux moins lourds mais stables, en g\'en\'eral bien au-del\`a du fer. C'est un processus de nucl\'eosynth\`ese connu, dit ``\textit{r}'' pour rapide, qui se produit dans un environnement riche en neutrons. Dans la kilonova, l'\'energie \'emise provient de la d\'esint\'egration radioactive des noyaux, et est environ 1000 fois sup\'erieure \`a celle d'une nova (d'o\`u son nom). Il est admirable de voir que ce mod\`ele a maintenant \'et\'e confront\'e avec succ\`es aux observations. L'analyse du spectre de la kilonova met en \'evidence des raies d'absorption caract\'eristiques des \'el\'ements lourds produits par le processus \textit{r}, voir la figure~\ref{fig10}.

\begin{figure}[h]
\begin{center}
\includegraphics[width=12cm]{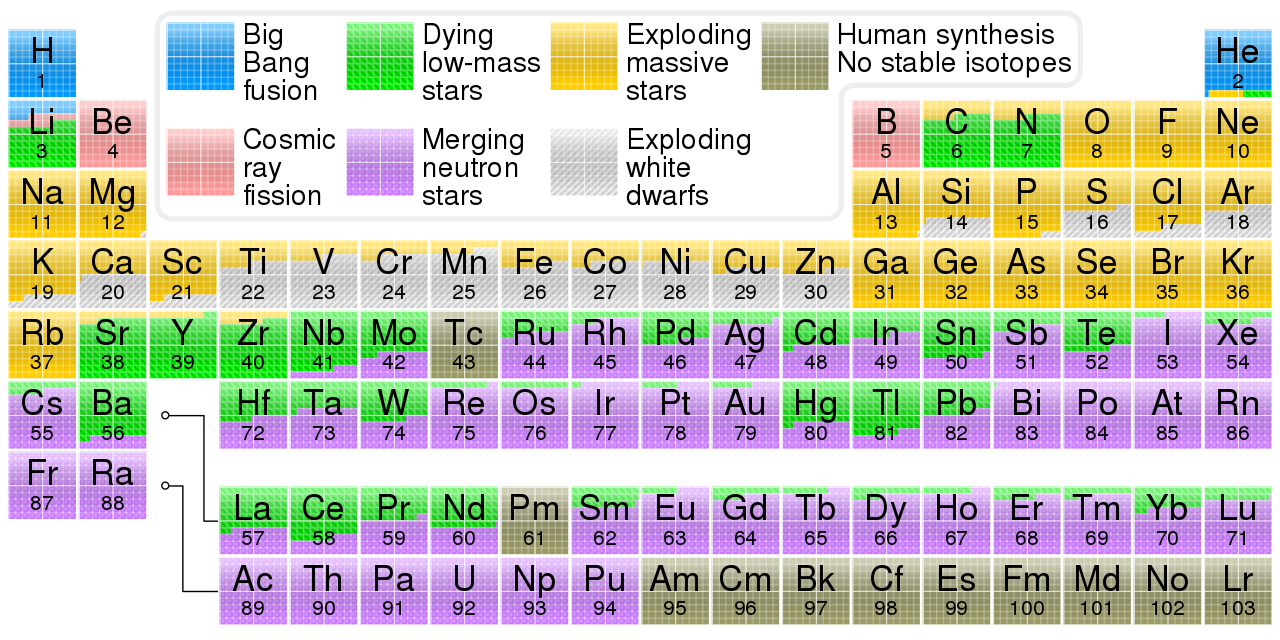}
\end{center}
\caption{La table des \'el\'ements avec leur origine probable. Les \'el\'ements les plus l\'egers tels que l'h\'elium sont synth\'etis\'es dans le Big-Bang. Pour beaucoup d'\'el\'ements plus lourds comme le fer, on invoque divers m\'ecanismes d'explosions de supernovas. La formation des \'el\'ements les plus lourds comme le c\'esium, l'or, les lanthanides, l'uranium, etc. est maintenant expliqu\'ee par les coalescences d'\'etoiles \`a neutrons (\textit{cf.} le code-couleur violet).}\label{fig11}
\end{figure}

La kilonova synth\'etise donc des noyaux tr\`es lourds, dont on avait du mal auparavant \`a expliquer l'origine. C'est donc une d\'ecouverte majeure et la r\'eponse \`a une question vieille de plus de 50 ans: les coalescences d'\'etoiles \`a neutrons et l'explosion cataclysmique associ\'ee constituent un site important de production d'\'el\'ements tels que l'or, le platine, l'uranium. La mod\'elisation d\'etaill\'ee des r\'eactions nucl\'eaires dans la kilonova et la confrontation aux observations permet de valider ce mod\`ele, et d'\'etudier les proportions d'\'el\'ements lourds qui sont synth\'etis\'es lors des coalescences d'\'etoiles \`a neutrons. Voir la figure~\ref{fig11}. 

\section{Tests de physique fondamentale} 

L'\'ev\'enement de fusion d'\'etoiles \`a neutrons GW170817 a d\'emontr\'e que la vitesse de propagation des ondes gravitationnelle est \'egale \`a celle de la lumi\`ere, \`a moins que $10^{-15}$ pr\`es. Ce simple fait est int\'eressant car il permet d'\'eliminer certaines th\'eories de la gravitation, autres que la relativit\'e g\'en\'erale. En particulier, des th\'eories qui ont \'et\'e propos\'ees comme alternatives \`a l'\'energie noire, c'est-\`a-dire pouvant expliquer l'acc\'el\'eration observ\'ee de l'expansion de l'Univers sans invoquer la constante cosmologique. Ces th\'eories \'evitent donc le fameux probl\`eme de la constante cosmologique, dont la valeur minuscule observ\'ee est difficilement explicable dans le cadre du mod\`ele standard de la cosmologie et de la th\'eorie quantique des champs~\cite{Martin12}. Ainsi, la classe de th\'eories ``tenseur-scalaire'' dite de Horndeski~\cite{Horndeski}, dans lesquelles un degr\'e de libert\'e scalaire est coupl\'e de fa\c{c}on g\'en\'erale au champ gravitationnel tensoriel de la relativit\'e g\'en\'erale, a-t-elle \'et\'e r\'ecemment fortement contrainte~\cite{CV17}.

La co\"incidence des temps d'arriv\'ee des signaux gravitationnel et \'electromagn\'etique (apr\`es un voyage de 130 millions d'ann\'ees), donne aussi un test du principe d'\'equivalence. En effet la trajectoire parcourue par les deux types de signaux dans l'Univers est la m\^eme. Or cette trajectoire est d\'evi\'ee par les grandes structures de l'Univers --- amas de galaxies, le super-amas local, le grand attracteur, \textit{etc.}, ce qui d\'emontre que diff\'erentes formes d'\'energies, \'electromagn\'etique et gravitationnelle, acqui\`erent la m\^eme acc\'el\'eration dans un champ gravitationnel. C'est exactement ce que l'on entend par principe d'\'equivalence. De plus, puisque l'une de ces \'energies est l'\'energie gravitationnelle, on a test\'e la version dite ``forte'' du principe d'\'equivalence.

Une autre avanc\'ee importante concerne la mesure de param\`etres cosmologiques. Cette possibilit\'e vient d'une propri\'et\'e remarquable des ondes gravitationnelles: elles contiennent en elles-m\^emes l'information sur la distance de la source, ici 40 Mpc. La mesure de cette distance, combin\'ee avec celle du d\'ecalage spectral (le ``redshift'' cosmologique) obtenue gr\^ace \`a la contrepartie \'electromagn\'etique, conduit \`a une estimation directe du param\`etre de Hubble $H_0$, qui d\'etermine le taux local d'expansion de l'Univers~\cite{Schutz86}. Dans le cas pr\'esent, avec une vitesse de r\'ecession de la galaxie NGC4993 de environ 3000 km/s, on obtient une valeur de $H_0$ de environ 70 km/s/Mpc, en accord avec d'autres estimations disponibles.

Pour finir, mentionnons un autre test int\'eressant, concernant la masse de la particule associ\'ee au champ de gravitation: le graviton. Cette masse est nulle en relativit\'e g\'en\'erale, mais elle a \'et\'e contrainte par les observations d'ondes gravitationnelles \`a \^etre inf\'erieure \`a $10^{-22}$ eV~\cite{LIGOtestGR}\,! L'id\'ee est que si le graviton est massif, sa vitesse d\'epend de son \'energie qui est elle-m\^eme proportionnelle \`a la fr\'equence, selon la fameuse formule de Planck $E = h \nu$. Or la fr\'equence n'est rien d'autre, \`a un facteur 2 pr\`es, que la fr\'equence orbitale du syst\`eme binaire, qui varie au cours du temps puisque c'est le chirp\,! On pr\'evoit donc dans le cas d'un graviton massif une distorsion du signal par rapport \`a la pr\'ediction de la relativit\'e g\'en\'erale~\cite{Will98}. Aucune distorsion n'ayant \'et\'e observ\'ee on en d\'eduit une valeur sup\'erieure pour la masse du graviton.

\bibliography{ListeRef_Dunod.bib}

\end{document}